\documentclass[a4paper,11pt]{article}
\usepackage{amssymb}
\usepackage{newtxtext,newtxmath}

\usepackage[T1]{fontenc}
\usepackage{rotating}
\usepackage{booktabs,multirow}
\usepackage{graphicx}
\usepackage{dcolumn}
\usepackage{bm}
\usepackage{subcaption} 
\usepackage{float}
\usepackage{caption}
\DeclareRobustCommand{\VAN}[3]{#2}
\let\VANthebibliography\thebibliography
\def\thebibliography{\DeclareRobustCommand{\VAN}[3]{##3}\VANthebibliography}

\usepackage{graphicx}	
\usepackage{amsmath}	

\usepackage[utf8]{inputenc}   
\usepackage{amsmath}          
\usepackage{graphicx}         
\usepackage{color}
\usepackage{hyperref}         
\usepackage{geometry}         
\usepackage{natbib}           
\usepackage{authblk}          
\usepackage{appendix}

\title{A continuous transition from Type-C Quasi Periodic Oscillations to the Heartbeat state in the Black hole X-ray binary 4U 1630--47}
\author[1]{Chintan Patel}
\author[2]{Sayantan Bhattacharya\thanks{
sayantan34@gmail.com}}
\author[1]{Karan Akbari}
\author[2]{Sudip Bhattacharyya\thanks{
sudip@tifr.res.in}}
\author[1]{Manojendu Choudhury}
\affil[1]{Department of Physics, St. Xavier's College, Mumbai, India, 400001}
\affil[2]{Department of Astronomy and Astrophysics, Tata Institute of Fundamental Research, 1 Homi Bhabha Road, Colaba, Mumbai 400005, India}
\date{\today}

\begin{document}

\maketitle

\begin{abstract}
We present a timing analysis of the black hole X-ray binary (BHXRB) 4U 1630–47 using \textit{AstroSat} observations from 10–19 March 2023, for the first time capturing a rare and rapid transition in variability properties. Within less than a day, the source evolved from a type-C quasi-periodic oscillation (QPO) state, with centroid frequencies between 3–5 Hz, to the Heartbeat state, characterized by a broad peak in the power density spectrum at $\sim$25 mHz, corresponding to a $\sim$40 s modulation period. 
As the source evolved, it passed through a transition track where the QPO features weakened and ultimately disappeared in the Heartbeat state. In the hardness-intensity Diagram, the QPOs occur at higher hardness and lower intensity, followed by a brightening phase as the source moved towards the soft intermediate state, and finally reached the Heartbeat state through a transition towards lower hardness. In the power-color diagram, this transition is marked by a clear shift to a distinct region of power color space, separate from the range occupied by other observed states.  This work establishes 4U 1630–47 as another system, apart from GRS 1915+105, where a continuous transition from QPO to Heartbeat state has been observed. Notably, 4U 1630–47 is the only system where the QPO is absent during the heartbeat state. This provides us with another probe to understand the physical mechanism governing this transition and the overall accretion mechanism in BHXRBs. 
\end{abstract}

\twocolumn
\section{Introduction}

Black hole X-ray binaries (BHXRBs) are systems in which a central stellar-mass black hole accretes matter from the companion star. 
These systems go through outburst phases caused by the rise of the accretion rate onto the black hole.
Throughout the outburst phase, the BHXRBs show significant evolution in their spectral and timing properties \citep{1994ApJS...92..511V}, based on which they are divided into different spectral states \citep{Homan_2005},  viz. the low-hard state, the hard intermediate state (HIMS), soft intermediate state (SIMS), and the high-soft state. These states can be traced on the hardness intensity diagram (HID) \citep{Homan_2005}, the power color diagram (PCD) \citep{10.1093/mnras/stv191}, and the hardness-rms diagram (HRD) \citep{10.1093/mnras/stu867}. \\

Timing analyses allow us to investigate the innermost region of the accretion flow \citep{Ingram_2019}. Low-frequency quasi-periodic oscillations (LFQPOs), with a centroid frequency in the range of 0.1-30 Hz, that are commonly observed in BHXRBs, are divided into types A, B, and C \citep{Casella_2005} with different spectral states linked to different types of quasi-periodic oscillations (QPOs) \citep{Ingram_2019}.\\

Few of the BHXRBs exhibit a state characterized by the light curve showing `electrocardiogram-like' oscillations having a low frequency modulation in the range of 10-20 mHz \citep{Neilsen_2011}, which result in a broad peak in the power density spectrum (PDS). This broad peak exhibits lower coherence and occurs at a significantly lower frequency and hence differs from typical QPOs which are characterized by narrow, high-coherence peaks. This state is generally referred to as quasi-regular modulation (QRM) \citep{2001MNRAS.322..309T, Zhao},  $\rho-$class \citep{2000A&A...355..271B}, and also the Heartbeat state \citep{Fan_2025}. The BHXRBS in which this Heartbeat state has been detected include GRS 1915+015 \citep{2000A&A...355..271B}, IGR J17091–-3624 \citep{Altamirano_2011,2021MNRAS.501.6123K}, GRO J1655-40 \citep{1999ApJ...522..397R} and 4U 1630–47 \citep{2001MNRAS.322..309T, Yang_2022}.

4U 1630--47 is a recurrent transient discovered by the Uhuru satellite \citep{1976ApJ...210L...9J} that goes into outburst every few hundred days \citep{Choudhury_2015}. Previous studies of the source during the 1998 outburst \citep{2001MNRAS.322..309T, Zhao} and the 2021 outburst \citep{Yang_2022} report the detection of quasi-regular modulation (QRM) and type C quasi-periodic oscillations (QPOs) existing independently or simultaneously without any direct causal connection between the QPO and the QRM / Heartbeat state.  In this work, we present \textit{AstroSat} LAXPC observations that capture for the first time a continuous transition in which type-C QPOs evolve along a distinct spectral-timing variability track and reach the QRM / Heartbeat state within the span of one day.

\begin{figure*}
    \centering

\begin{subfigure}[b]{0.8\textwidth}
    \centering
    \includegraphics[width=\linewidth]{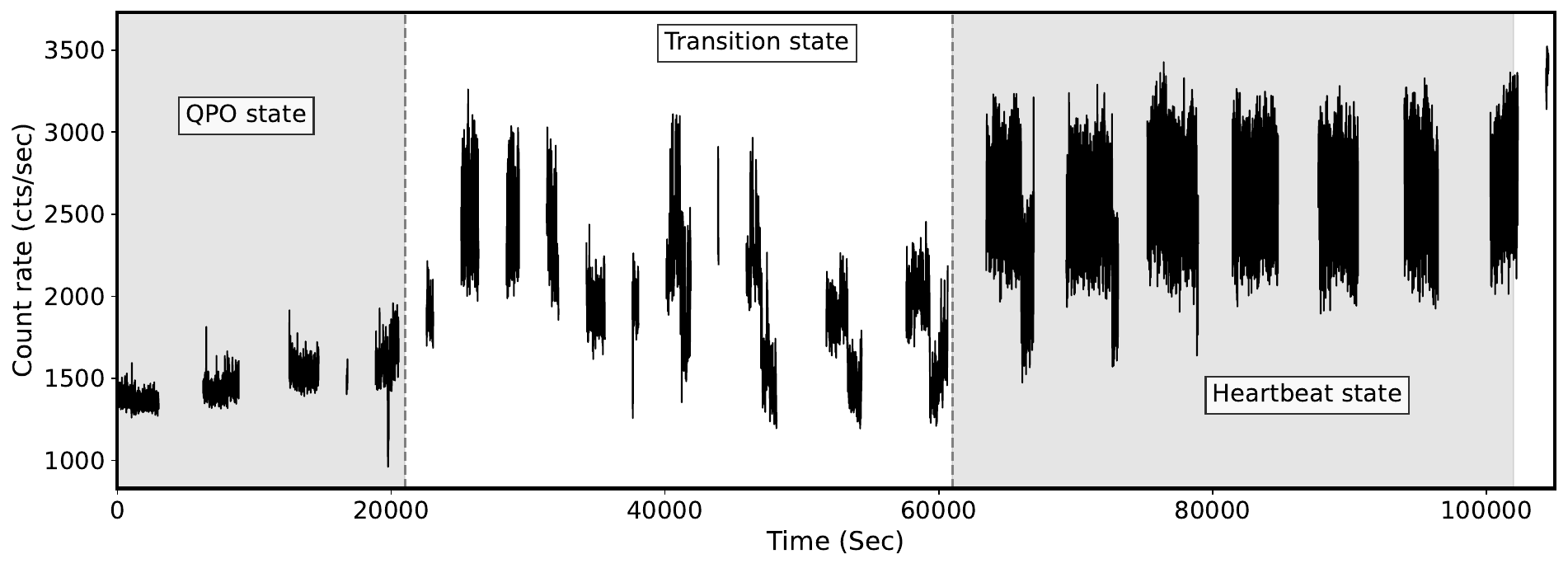}
\end{subfigure}
\hfill

    \begin{subfigure}[b]{0.32\textwidth}
        \centering
        \includegraphics[width=\linewidth]{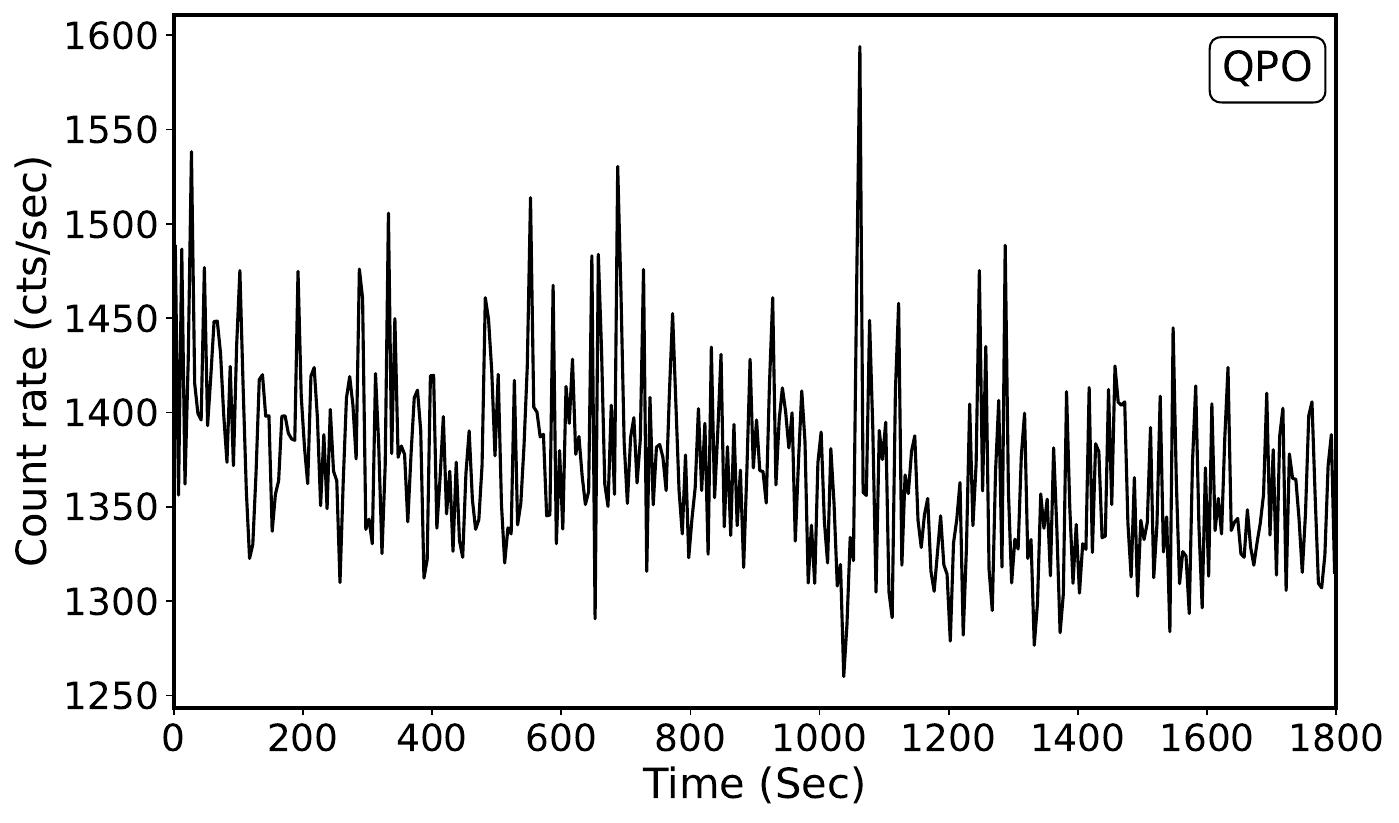}
    \end{subfigure}
    \hfill
    \begin{subfigure}[b]{0.32\textwidth}
        \centering
        \includegraphics[width=\linewidth]{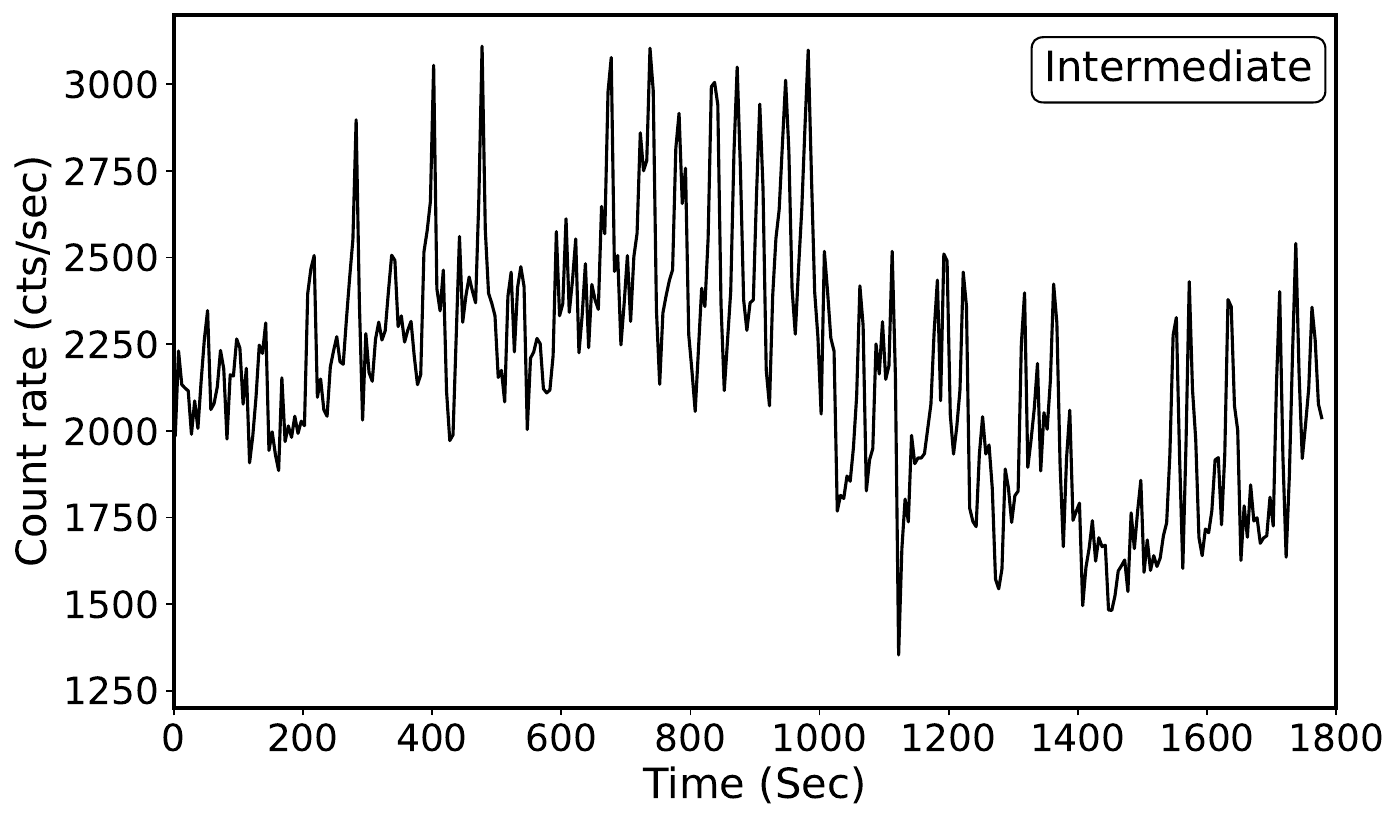}
    \end{subfigure}
    \hfill
    \begin{subfigure}[b]{0.32\textwidth}
        \centering
        \includegraphics[width=\linewidth]{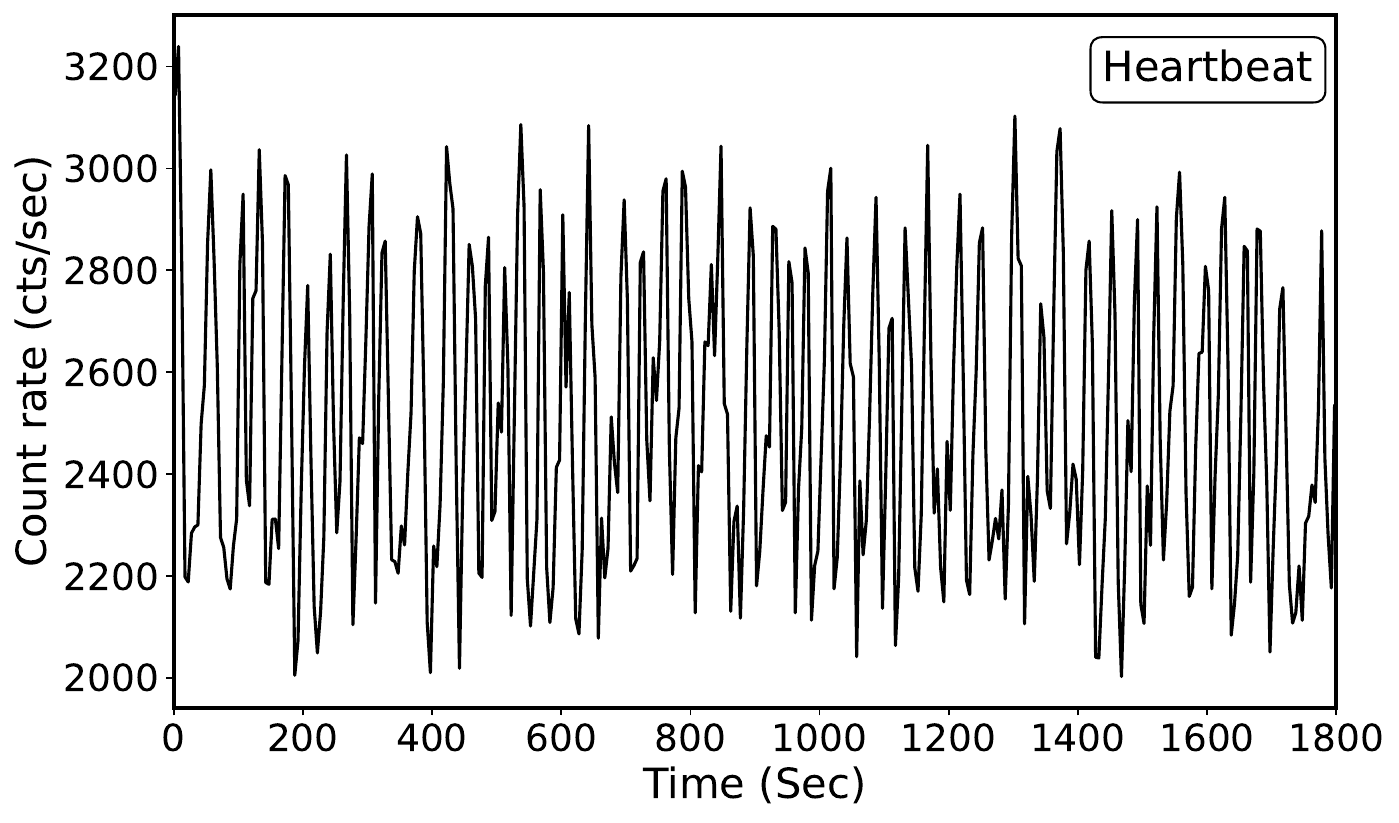}
    \end{subfigure}

    \vspace{1em}

    \begin{subfigure}[b]{0.32\textwidth}
        \centering
        \includegraphics[width=\linewidth]{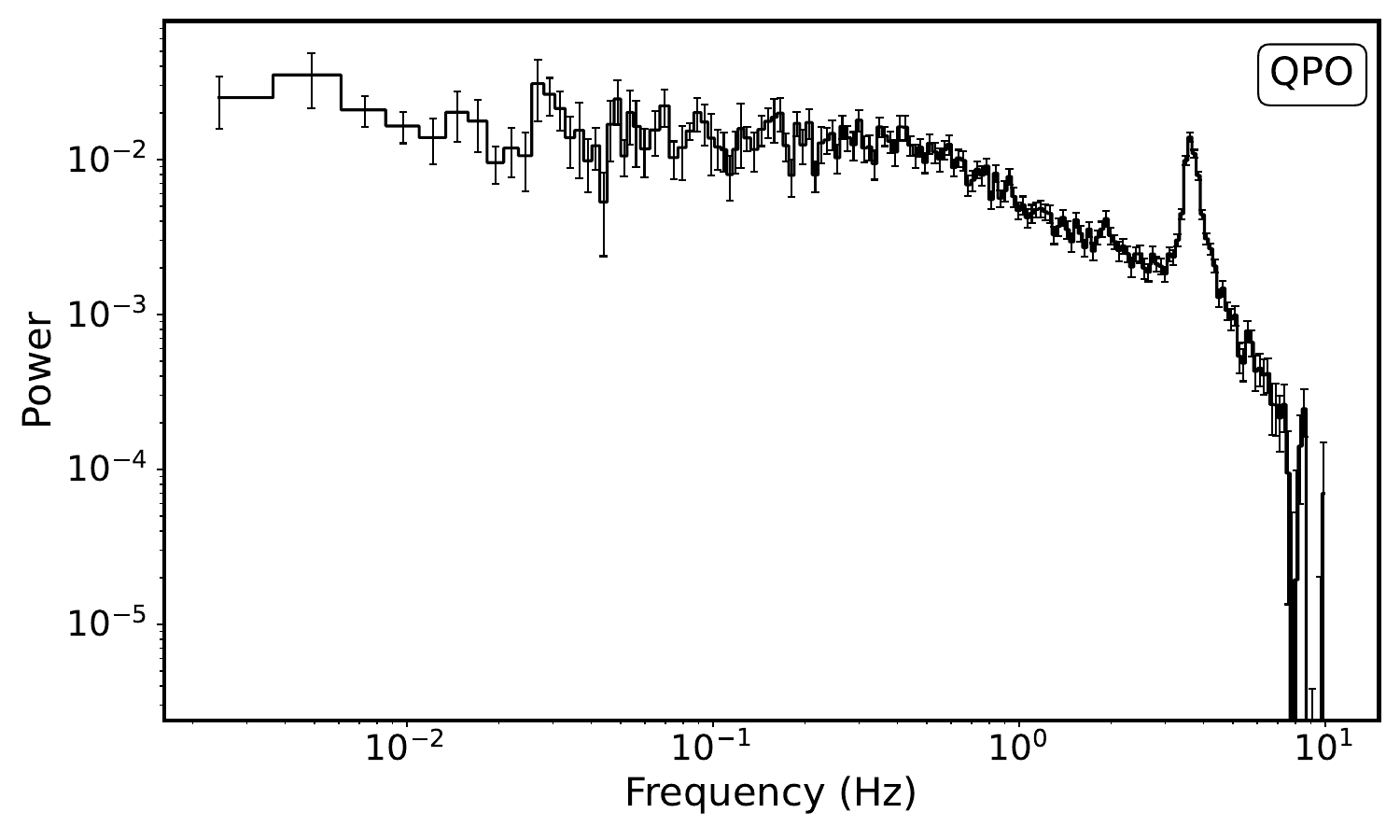}
    \end{subfigure}
    \hfill
    \begin{subfigure}[b]{0.32\textwidth}
        \centering
        \includegraphics[width=\linewidth]{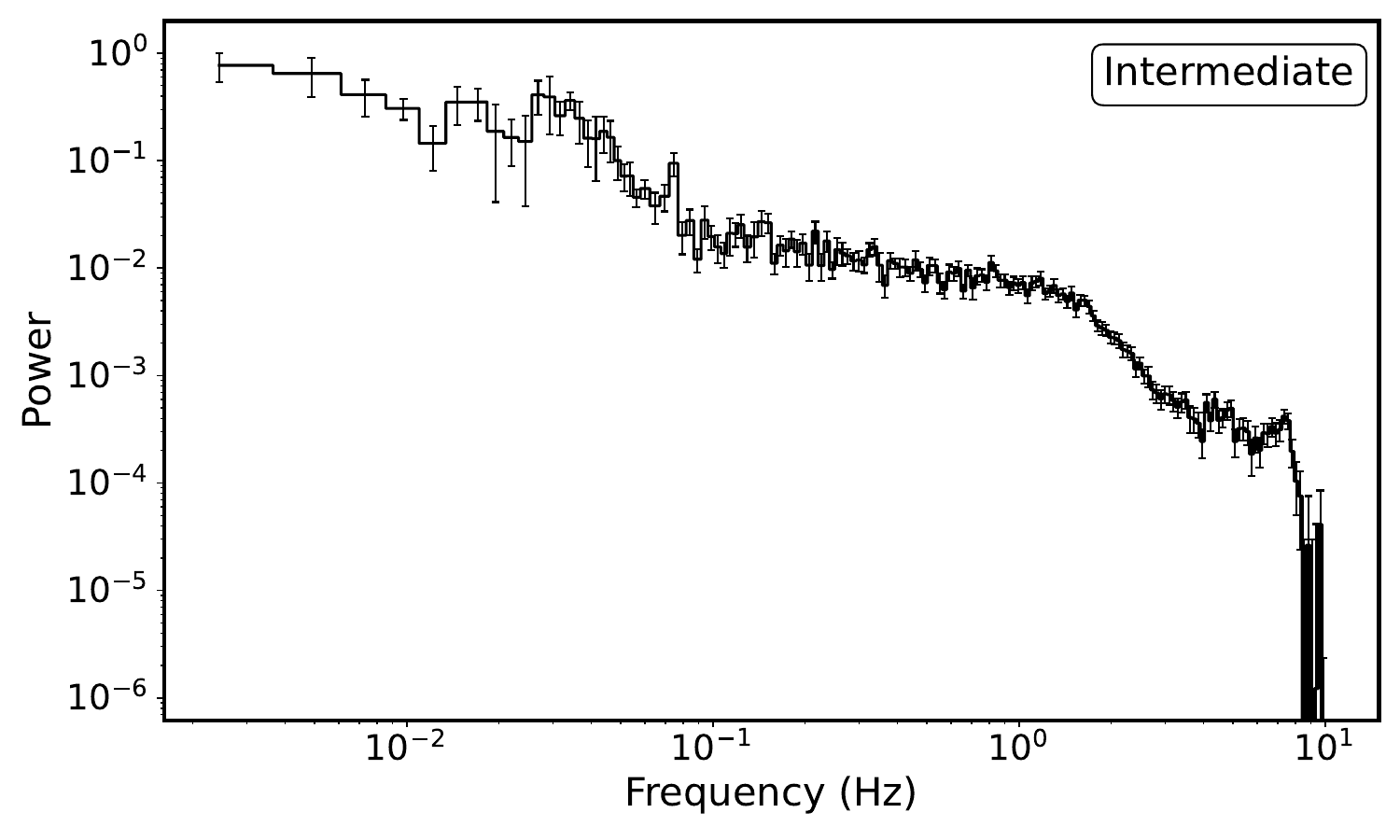}
    \end{subfigure}
    \hfill
    \begin{subfigure}[b]{0.32\textwidth}
        \centering
        \includegraphics[width=\linewidth]{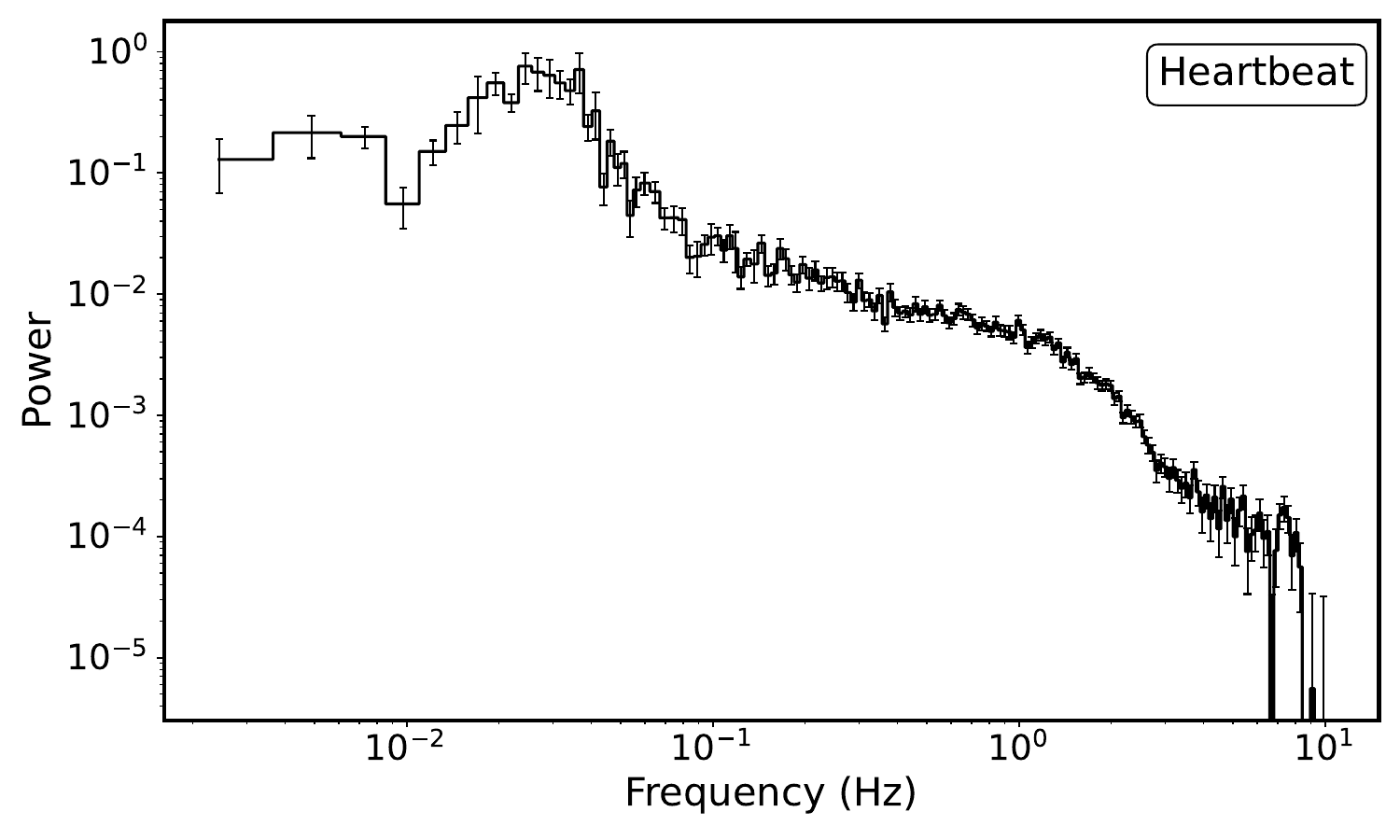}
    \end{subfigure}

    \caption{
        \textit{Top row:} Overall summary of variability evolution in 4U 1630–47 as observed on 15 March 2023 (left) and 16 March 2023 (right) observed with \textit{AstroSat} LAXPC. The zero of the time-axis corresponds to MJD 60018.0284 and ends at MJD 60019.1891. \textit{Middle and bottom row:} representational segments of 1000s of the LAXPC light curves (middle) from each variability state and corresponding power density spectra (bottom row) for three distinct states — \textit{Left:} Type-C QPO, \textit{Middle:} transition state, and \textit{Right:} Heartbeat state.
    }
    \label{fig1}
\end{figure*}

\section{Observation \& Data Analysis} \label{sec:DA}

The source 4U 1630--47 was observed by \textit{AstroSat} between 10 and 19 March 2023: initially during 10--12 March 2023 followed by a continuous stretch of observations from 15 to 19 March 2023. These observations were part of a Target of Opportunity (TOO) campaign. Here we report the analysis of the observation ID \texttt{A12\_056T02\_9000005538}\footnote{\url{https://astrobrowse.issdc.gov.in/astro_archive/archive/Home.jsp}} covering 74 orbits with an exposure total exposure time of 716305.90 s. The observations by the Large Area X-ray Proportional Counter (LAXPC) aboard the \textit{AstroSat} were conducted in the event mode. In this work the data from the unit LAXPC20 have been analyzed for its overall consistent performance through the lifespan of \textit{AstroSat}. The data was reduced using \texttt{LaxpcSoftv3.4.3}\footnote{\url{https://www.tifr.res.in/~astrosat_laxpc/software.html}} developed by the Tata Institute of Fundamental Research. The level 1 data is acquired from the \textit{AstroSat} archive and is processed to obtain the level 2 data from which the final products are generated. The data extraction processes were performed using format A\footnote{\url{http://astrosat-ssc.iucaa.in/uploads/threadsPageNew_SXT.html}} adhering to the processes laid down by the \textit{AstroSat} science support cell. 
The timing analysis and other processes related to X-ray astronomical data analysis were conducted using \texttt{HEASOFT v6.35}\footnote{\url{https://heasarc.gsfc.nasa.gov/docs/software/lheasoft/}}. \\

The power density spectra (PDS) were obtained for each segment of the LAXPC light curves during 10--19 March for detailed investigation. The overall spectral evolution vis-a-vis the timing analysis during this whole window of observation was achieved by scrutinizing the hardness intensity diagrams (HID). For this paper the hardness ratio is defined by the energy bands 7--40 keV and 3--7 keV. A more detailed investigation of the physical processes was undertaken by obtaining the power color diagram (PCD) by plotting the ratios of integrated variability power in different frequency bands \citep{10.1093/mnras/stv191} over the whole stretch  of 10--19 March. The PDS were divided into three frequency bands: 0.0024–0.1 Hz, 0.1–1 Hz, and 1–10 Hz (band1, band2, and band 3 respectively). The variance in each band was computed by integrating the PDS, with the Poisson noise contribution retained as the relatively low power observed on several days were leading to negative variance if the noise was excluded. Error propagation was carried out following the standard procedure outlined by \citet{vanderKlis1989}. The relationship between spectral hardness and fractional rms variability is examined by the hardness rms diagram (HRD) \citep{10.1093/mnras/stu867}. The fractional rms is computed by integrating the power density spectrum (PDS) over a selected frequency range and taking the square root of the integrated power.\\

\section{Results}\label{sec:res}
The light curve during 15–16 March 2023 (MJD 60018–60019), shown in the top panel of Figure \ref{fig1}, reveals that the spectral-timing evolution is not solely flux-dependent. Based on the variability properties in the PDS, the light curve is divided into three variability class: QPO, transition, and Heartbeat. Due to a gap in observations from 12–15 March, the onset of the QPO state is not captured. For 16 March (MJD 60019), only the Heartbeat segment is shown in the top panel. The middle row of Figure \ref{fig1} displays representative light curve segments for each state, while the bottom row presents their corresponding PDS.\\

\begin{figure*}
    \centering
    \includegraphics[width=0.95\textwidth]{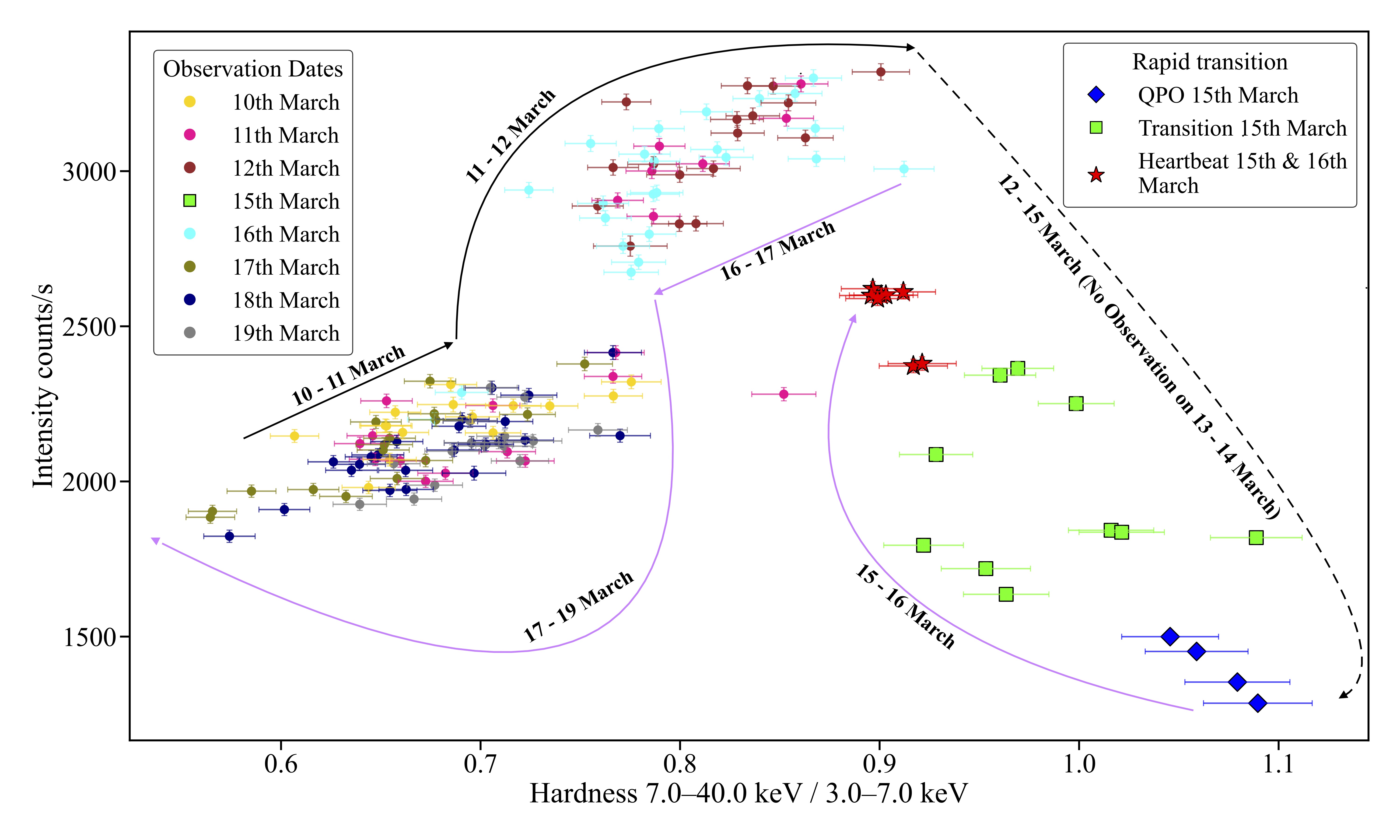}
    \caption{The HID of LAXPC observations of 4U 1630--47 from 10–19 March 2023. The hardness is defined as the ratio of count rates in the 7–40 keV and 3–7 keV bands. Observations are color coded as per the day of observation with different color and marker for the temporally instable states. The blue diamond represents the QPOs which occur on 15 March, and the red star corresponds to the Heartbeat state which occurs on 15 and 16 March (5 on 15 March and 3 on 16 March, as shown in Figure \ref{fig1}, top panel). The solid black arrow represents the transition from SIMS to HIMS and the dotted black line represents jump from 12 to 15 March (no observations on 13 \& 14 March). The purple arrow shows the reverse transition. }
    \label{fig2}
\end{figure*}

During MJD 60018.02840--60018.24666 (0-20000 s of figure \ref{fig1} top panel), the PDS (figure \ref{fig1} bottom row left panel) exhibits a characteristic flat-top noise at low frequencies (0.01-0.03 Hz), indicating a strong aperiodic variability likely due to instabilities in the outer accretion disk. A clear break is observed near 0.5 Hz, marking the transition from flat-top to steeply declining power, often interpreted as the dynamic timescale of the inner accretion flow \citep{Ingram_2019}. Superimposed on this continuum is a prominent QPO, consistent with the picture of type-C QPO observed in hard intermediate state of BHXRBs \citep{Casella_2005}. This segment is classified as the QPO state.\\
During MJD 60018.28970--60018.69531 (20000-60000 s of figure \ref{fig1} top panel), the PDS (figure \ref{fig1} bottom row middle panel) shows that the broadband noise component is reduced and the QPO feature is very weak, likely due to onset of transition in the X-ray state, causing the QPO to diminish. Here the PDS displays a steeply declining power-law-like behavior from low to high frequencies, with low fractional rms variability. Thus, we can infer that the system is undergoing a transition towards a soft intermediate state.\\
During MJD 60018.76292--60019.18914 (60000-90000 s of figure \ref{fig1} top panel), the PDS (figure \ref{fig1} bottom row right panel) is characterized by a prominent broad peak at $\sim$25 mHz and displaying a smooth and featureless continuum, while the corresponding light curve (figure \ref{fig1} middle row right panel) shows regular flares and dips in approximately 40 seconds. \\

\begin{figure*}
    \centering
    \includegraphics[width=0.95\textwidth]{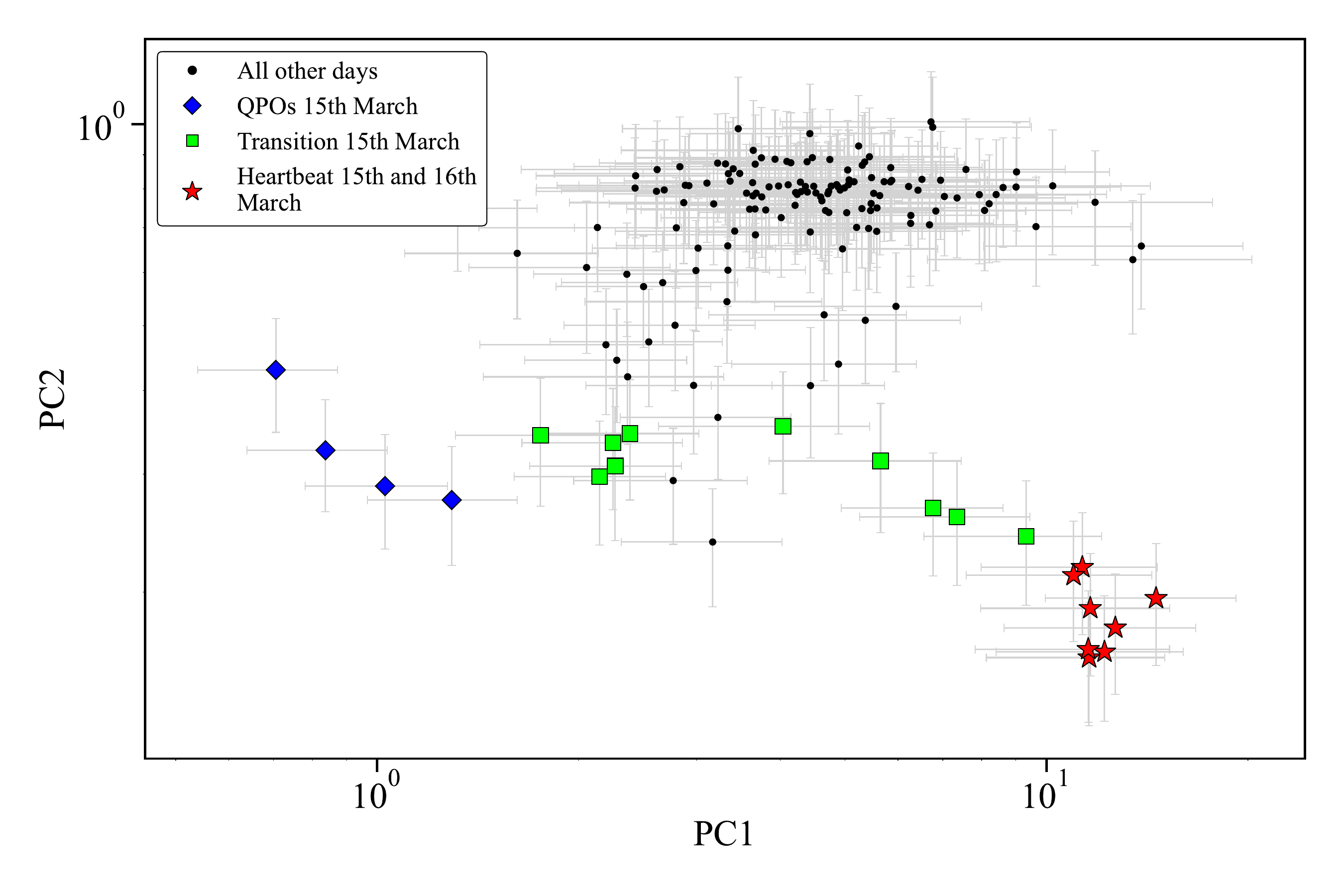}
    \caption{ The PCD \citep{10.1093/mnras/stv191} where the PDS is divided into three frequency bands: band1 (0.0024–0.1 Hz), band2 (0.1–1 Hz), and band3 (1–10 Hz). PC1 is the ratio of the variance band1/band2, and PC2 is band3/band2.}
    \label{fig3}
\end{figure*}

Figure~\ref{fig2} presents the HID of 4U 1630–47 during the period 10–19 March 2023. The data points are color-coded by observation date to highlight the spectral evolution. The source covers a closed loop in the parameter space starting with the stable phase (i.e. relatively stable timing features) at relatively low flux and low hardness ratio to a state with higher flux and greater hardness ratio when the source goes into the instability phase (variability in timing features), following which the source gets back to the previous stable phase with similar flux and hardness ratio before falling back to the initial state.   \\

The beginning (10 March) and the end (17, 18 \& 19 March) of the whole episode has the source exhibiting moderate values of hardness ratio ranging from approximately 0.65 to 0.80, with the corresponding photon flux remaining relatively steady between $\sim$2000 and 2300 counts s$^{-1}$. Interestingly, the source evolves away and falls back to this state in a discontinuous manner to and from a state characterized by values of hardness ratio of up to $\sim$0.9 and photon flux peaking at nearly 3500 counts s$^{-1}$ (on 11 \& 12 March, and on 16 \& 17 March, respectively). In between (during 15 -- 16 March), the source evolves through the instability phase. This phase is characterized by (1) the LFQPO phase where the source displays a noticeable drop in intensity to $\sim$1000 counts s$^{-1}$. It also shows a substantial hardening, with a hardness ratio in the range of 1.0--1.1. This is followed by (2) the transition phase characterized by hardness ratio softening from $\sim$1.1 to 0.9 and the photon counts increasing to $\sim$2500 counts s$^{-1}$, followed by 3) the Heartbeat phase, with the hardness ratio around $\sim$0.9 and the photon count increasing above $\sim$2500 counts s$^{-1}$. \\

As shown in Figure \ref{fig3} bottom left panel, the PCD presents the evolution of the source’s timing properties \citep{10.1093/mnras/stv191}, where the x-axis corresponds to the variance ratio between the soft band and the intermediate band (band1/band2), while the y-axis represents the variance ratio between the hard band and the intermediate band (band3/band2). This technique provides a model-independent method to characterize the broadband power spectral shape and to distinguish between different accretion states in X-ray binaries. The observations of stable state (black markers) are clustered in a broadly similar region. However, during the instability phase (15--16 March) (blue, green, and red markers) the parameters reveal significant deviation, with clusters corresponding to regions associated with the QPOs, the transition, and the Heartbeat states, respectively. The values of power-color parameters of the Heartbeat region are indicative of strong, quasi-regular oscillations likely driven by radiation-pressure dominated instabilities in the inner accretion disk \citep{lightman1974}. Meanwhile, the same for QPOs  suggest the presences of localized precessing motion, commonly interpreted as evidence for frame-dragging of the inner accretion flow or corona \citep{Ingram_2019}. The sharp transition between these variability regimes on a single day implies rapid evolution of the inner accretion flow, potentially related to changes in mass accretion rate or the disk/corona geometry. Such fast transitions are characteristic of intermediate accretion states \citep{2011BASI...39..409B}. In the following subections, we quantitatively discuss the nature of the three staes. The PDS is fitted with three Lorentzian components with its centroid frequency $f_0$, full width at half maximum (FWHM), quality factor $Q=f_0/\mathrm{FWHM}$, signal-to-noise ratio (SNR), and fractional rms amplitude $r$.


 \subsection*{QPO State}
A single strong quasi-periodic oscillation (type-C QPO) appears at 
$f_{0} \approx 3.7$--$5.7$~Hz with high coherence ($Q \approx 4$--7) and large 
rms (up to $\sim 15\%$). A secondary hump component near 1--2 Hz is broad (low $Q$) initially weaker rms but it gets stronger  ($r \sim 7\%$--25\%). Over these segments the QPO centroid shifts upward slightly 3.7~Hz $\rightarrow$ 5.7~H) consistent with an increasing accretion rate or shrinking inner disk radius.  

\subsection*{Intermediate State}
The 5--6~Hz QPO is still present but steadily weakens. Its SNR and rms drop, and its $Q$-value declines, indicating reduced coherence. Crucially, a very-low-frequency component appears: a heartbeat peak at $\sim 0.01$--0.03~Hz ($\sim 10$--30~mHz) with moderate  $Q \sim 1$--2 and very high rms (up to $\sim 30\%$).  The hump remains at $\sim 1$~Hz but also weakens. This intermediate regime shows both the fading type-C QPO and the growing 
heartbeat modulation. It is reported that such QPO--QRM transitions can occur 
very rapidly. Fast recurrence of the mHz QRM, in a timescale of days, has been observed when the source transitions from the QRM state to the non-QRM state.
In our case, the full transition (QPO~$\rightarrow$~heartbeat) took on the order of less than a day.  

\subsection*{Heartbeat State}
Once fully in the heartbeat (QRM) regime, the original few-Hz QPO is 
essentially gone. The PDS is dominated by the broad, low-frequency modulation at 
$\sim 0.02$--0.03~Hz (period $\sim 20$--40~s) with high amplitude 
(rms $\sim 20$--30\%). The remaining hump noise is at very low frequency (around 0.1--1~Hz) with low significance. In this state, the light curve shows large quasi-periodic flares (an electrocardiogram-like pattern) every $\sim 30-40$~s, such lightcurves with different time modulations were reported in 4U~1630--47's heartbeat episodes \cite{Fan_2025}. \\

The evolution matches theoretical and observational models of state transitions in black-hole binaries. Initially, a hard-intermediate state produces a strong type-C QPO at a few Hz, linked to Lense--Thirring precession of the inner flow or corona \cite{stella_luigi} \cite{Ingram_2019}.As the accretion rate rises and the spectrum softens, this QPO weakens and disappears.  Simultaneously, the disk enters a radiation-pressure dominated regime, 
triggering a limit-cycle instability. This produces the low-frequency heartbeat oscillation in the light curve \cite{Yang_2022}. The heartbeat's $\sim 30-40$~s timescale is roughly the viscous time of the inner disk \cite{sheifina}, and its large fractional amplitude reflects a large instability-driven modulation in the disk emission. Observationally, past work on 4U~1630--47 \cite{Trudolyubov2001}. \citep{Yang_2022} similarly finds that the source transitions sharply from a few-Hz QPO to an mHz QRM.

\section{Discussion}
The analysis of the continuous observations of 4U 1630--47 establish the evolution of the source from a QPO to a Heartbeat state via a transition. The physical implications of these variabilities ultimately connect with the underlying mechanism of the said evolution.

\subsection{Type-C QPOs}
Type-C quasi-periodic oscillations (QPOs) are among the most frequently observed low-frequency QPOs in black hole X-ray binaries (BHXRBs), typically emerging during the hard and hard-intermediate spectral states. These QPOs are characterized by centroid frequencies ranging from approximately 0.01 to 30 Hz, high coherence ($Q \gtrsim 10$), and significant rms amplitudes often exceeding 10–15\%.\\

Various mechanisms have been proposed to theorize these temporal instabilities. The accretion–ejection instability (AEI) model suggests a magnetohydrodynamic spiral instability in the inner disk \citep{1999A&A...349.1003T, 2002A&A...387..497V}. However, its applicability to the full phenomenology of Type-C QPOs is limited by a lack of robust observational signatures. Other proposals, such as shock oscillation models within two-component accretion flows \citep{Chakrabarti_2000}, or intrinsic pressure or diskoseismic oscillations \citep{Wagoner:2001uj}, face challenges in accounting for the observed frequency ranges, inclination effects, and spectral–timing correlations.\\

The most widely accepted model attributes Type-C QPOs to Lense–Thirring precession of a geometrically thick, optically thin, hot inner flow that is misaligned with the black hole spin axis. In this picture, the quasi-periodic modulation of the X-ray lightcurve is primarily geometric: as the hot flow precesses its projected area and line-of-sight velocity change, producing periodic changes in the observed Comptonized flux through variations in projected solid angle, Doppler boosting, and the anisotropic emission pattern of the corona. Reflection from the disk is modulated as the illumination pattern changes, producing phase-dependent spectral signatures such as an oscillating iron-K profile \citep{2015MNRAS.446.3516I} and small changes in reflection fraction, but these reflected components usually make up only a modest fraction of the total variable flux \citep{1998ApJ...492L..59S, Ingram_2019, Ingram_Done_2011}.

\subsection{Heartbeat state}
The Heartbeat state is a unique and luminous oscillatory accretion regime observed primarily in the BHXRB GRS 1915+105 and, more recently, in IGR J17091–-3624. It shows a repetitive cycle of rapid rise and decay in X-ray luminosity on timescales of 10–100 seconds \citep{Belloni2000,Altamirano_2011}. These oscillations are believed to be driven by a limit-cycle instability in the inner accretion disk, wherein the disk alternates between high and low accretion states near or at the Eddington luminosity involving a radiation pressure instability operating in the innermost regions of a radiation-dominated accretion disk \citep{2000ApJ...542L..33J,Nayakshin2000}. When the local radiation pressure exceeds the viscous stress, the disk becomes thermally unstable, causing cyclic evacuation and refilling of the inner accretion zone. This process shows up observationally as periodic flares and dips, accompanied by rapid spectral transitions from hard to soft states. General relativistic magnetohydrodynamic (GRMHD) simulations incorporating radiation transport support the existence of such limit cycles under near-Eddington accretion rates \citep{2017MNRAS.466..705S, Mishra_2022}.\\


\subsection{The rapid transition of 4U 1630--47}
The evolution in the HID reveals the spectral characteristics of the accretion flow in 4U 1630–-47. The initial rise in both hardness and intensity on 11 March is indicative of increasing accretion rate, $\dot{M}$, as the source transitions deeper into the intermediate state. The high hardness values ($\sim$0.9) paired with increased flux ($>$3000 counts s$^{-1}$) are characteristic of a HIMS, in which a hot, optically thin inner flow coexists with a truncated outer accretion disk \citep{Homan_2005, Done2007}. The disk is yet to reach near innermost stable circular orbit (ISCO), and variability is governed by mechanisms such as Lense–Thirring precession, giving rise to type-C QPOs \citep{Ingram_2019}.\\

The sharp change observed on 15 March, when the source appears at a significantly lower intensity ($\sim$1000 counts s$^{-1}$) and higher hardness ($>1.0$), is unusual and may be interpreted as a brief drop in accretion rate or a change of geometry of the inner flow, possibly due to the temporary retreat of the inner disk radius. The simultaneous appearance of strong QPOs during this phase indicates that a precessing hot inner flow exists. As the day progresses, the increasing intensity and decreasing hardness suggest that the accretion disk moves inward. This is supported by the disappearance of QPOs and the emergence of large-amplitude, low-frequency Heartbeat oscillations in the light curve.\\


The return path observed during 16–19 March, where the source retraces its position back to the initial state suggests a reduction in $\dot{M}$ below the instability threshold. This results in the stabilization of the inner disk, causing disappearance of Heartbeat feature, and restoration of a stable, geometrically thin accretion regime. The loop-like pattern formed in the HID is similar to hysteresis tracks observed in other black hole transients\citep{Homan_2005}.\\

During the given observational evolution of the source, the causal connection of the QPO and the Heartbeat states form one of the most interesting features in the accretion process of the stellar mass black hole binaries, where seamless transition between the seemingly disparate physical conditions of the physical conditions of the accretion disc/coronal plasma that exist in the QPO and the Heartbeat phases individually. The transition from a QPO-like scenario to a Heartbeat-like scenario implies that the accreting matter was first undergoing turbulence with the QPO characteristic frequency of 3.6 - 4.8 Hz and the emitted photons were in the hard band (see figure \ref{fig2}). 
The latter heartbeat scenario presents as a broad peak in the PDS and at a much lower frequency; this is a more stable (but slowly oscillating) feature emitting relatively softer photons. These QPO and heartbeat features appear one after another, but there is no strong evidence if one of them is actually causing another.\\

Ultimately, on 16 March, the PDS feature ceases to exist and the source is back to the previous stable phase, as indicated in both PCD and HID.  Hence, we can only speculate that the type-C QPO might be the precursor of the Heartbeat state. The return to stability may be a single cycle of a continuous hysteresis chain (as seen in later \textit{NICER} observation a week later \citep{Fan_2025}). This observation emphasizes the importance of continuous monitoring of these transitions (because pre- and post-Heartbeat are similar if observed with a gap) 
More such observations will determine if the frequency or spectral nature during the Heartbeat phase depends on the precursor QPO event, and if yes, then, to what extent?     

A similar transition was observed in GRS 1915 +105 \citep{2019ApJ...870....4R} where the source evolved from $\chi$-- class to Heartbeat state. Both the sources exhibit transitions from a type-C QPO state to a Heartbeat state, implicating similar mechanism, involving the radiation‐pressure instabilities in the inner disk \citep{lightman1974}. Observationally, both transitions involve source softening and the emergence of strong slow X-ray modulations with timescales set by the viscous timescale of the inner disk: 100–150s in GRS 1915+105 \cite{2019ApJ...870....4R}, $\sim$40s in 4U 1630--47, reflecting different radii, spin\citep{pahari} and BH masses.  In both cases the type-C QPO frequency increases slightly as the source softens, consistent with an inward-moving disk/corona, and then in 4U 1630-47 a rapid transition where the QPO weakens and diminishes gives way to the low‐frequency Heartbeat.  Nevertheless, the key differences between the sources are: i) GRS 1915+105 displays a looping CCD track during the Heartbeat \citep{2019ApJ...870....4R}, whereas the HID 4U 1630–-47 shows a pronounced hysteresis loop and a distinct jump in PCD space during the transition,  and ii) in GRS 1915+105 the QPO persists (though weakened) in both the intermediate and  Heartbeat state, whereas in 4U 1630–47 the QPO weakens (not coherent enough to qualify as QPO) during the transition state (intermediate state) and ultimately disappears in Heartbeat state , implying a more abrupt structural change.

Physically, these differences suggest that while both systems undergo radiation‐pressure limit cycles, the trigger and timescale of the same differ.  GRS 1915+105 is nearly always at high accretion, so its Heartbeat may represent the precursor of a full limit‐cycle; its gradual loop in CCD hints at a more gradual change in the coronal geometry.  In contrast, 4U 1630–-47 underwent a fast outburst rise through the HIMS, reaching a tipping point where RPI took over very rapidly, then returning to a stable disk state.  

\section{\textbf{Conclusion}}
In this work, we present the spectral and timing evolution of the black hole X-ray binary 4U 1630–47, using \textit{AstroSat} LAXPC data from its 2023 outburst. This marks the first clear observational evidence of a rapid ($>$ 24 hours), continuous transition from a type-C QPO state to the Heartbeat or QRM state. Through detailed analysis, involving PDS, HID, PCD, and HRD, we trace the evolution of the system’s characteristics before, after and across this aforementioned transition.
Here are the main points.

\begin{enumerate}
    \item We observe in the PDS the appearance of a QPO feature on 15 March 2023, with a centroid frequency of 3.7-4.8 Hz. Then, an intermediate state was observed when the source was re-brightening and getting softer, but exhibited no prominent features in the PDS. Towards the end of 15 March 2023, observations of a Heartbeat-like feature with a centroid frequency of 25-28 mHz are observed in the PDS and sustained till the initial three light curve segments of 16 March 2023. After that, the source transitions back to its initial state in all of HID, PCD and HRD.
   \item The observed LFQPOs could originate due to a Lense-Thirring precession of the hot inner flow. Then the source transits towards a softer state, suggesting a decreased coronal size. This causes the fast inner oscillations to spread in a larger region and get more regular but slower. This phenomenon is observed in the PDS as a Heartbeat-like feature.
\end{enumerate}
The results of this study shows the Heartbeat feature which follows a pre-existing type-C QPO, and establishes a possibility that both of these can exist in a loop within a specific hardness-intensity regime. To develop an overall understanding of what scenarios can cause a such loop of QPO to a Heartbeat transition, more continuous and longer duration monitoring observations are required.

\section{acknowledgments}
This research has made use of data obtained from the {\it \textit{AstroSat}} mission of the Indian Space Research Organization (ISRO), archived at the Indian Space Science Data Center (ISSDC). We thank the SXT and LAXPC Payload Operation Centers at TIFR, Mumbai, for verifying and releasing the data. We also thank the respective instrument teams for their assistance with the data reduction and analysis. This research has made use of software provided by the High Energy Astrophysics Science Archive Research Center (HEASARC), which is a service of the Astrophysics Science Division at NASA/GSFC.

\section*{Data Availability}

The data used is publicly available through \textit{AstroSat} archive.

\bibliographystyle{mnras}
\bibliography{ref} 

\begin{thebibliography}{}
\makeatletter
\relax
\def\mn@urlcharsother{\let\do\@makeother \do\$\do\&\do\#\do\^\do\_\do\%\do\~}
\def\mn@doi{\begingroup\mn@urlcharsother \@ifnextchar [ {\mn@doi@} {\mn@doi@[]}}
\def\mn@doi@[#1]#2{\def\@tempa{#1}\ifx\@tempa\@empty \href {http://dx.doi.org/#2} {doi:#2}\else \href {http://dx.doi.org/#2} {#1}\fi \endgroup}
\def\mn@eprint#1#2{\mn@eprint@#1:#2::\@nil}
\def\mn@eprint@arXiv#1{\href {http://arxiv.org/abs/#1} {{\tt arXiv:#1}}}
\def\mn@eprint@dblp#1{\href {http://dblp.uni-trier.de/rec/bibtex/#1.xml} {dblp:#1}}
\def\mn@eprint@#1:#2:#3:#4\@nil{\def\@tempa {#1}\def\@tempb {#2}\def\@tempc {#3}\ifx \@tempc \@empty \let \@tempc \@tempb \let \@tempb \@tempa \fi \ifx \@tempb \@empty \def\@tempb {arXiv}\fi \@ifundefined {mn@eprint@\@tempb}{\@tempb:\@tempc}{\expandafter \expandafter \csname mn@eprint@\@tempb\endcsname \expandafter{\@tempc}}}

\bibitem[\protect\citeauthoryear{Altamirano et~al.,}{Altamirano et~al.}{2011}]{Altamirano_2011}
Altamirano D.,  et~al., 2011, \mn@doi [The Astrophysical Journal] {10.1088/2041-8205/742/2/l17}, 742, L17

\bibitem[\protect\citeauthoryear{{Belloni}, {Klein-Wolt}, {M{\'e}ndez}, {van der Klis}  \& {van Paradijs}}{{Belloni} et~al.}{2000a}]{2000A&A...355..271B}
{Belloni} T.,  {Klein-Wolt} M.,  {M{\'e}ndez} M.,  {van der Klis} M.,   {van Paradijs} J.,  2000a, \mn@doi [AA] {10.48550/arXiv.astro-ph/0001103}, \href {https://ui.adsabs.harvard.edu/abs/2000A&A...355..271B} {355, 271}

\bibitem[\protect\citeauthoryear{Belloni, Klein-Wolt, Mendez, van~der Klis  \& van Paradijs}{Belloni et~al.}{2000b}]{Belloni2000}
Belloni T.,  Klein-Wolt M.,  Mendez M.,  van~der Klis M.,   van Paradijs J.,  2000b, A\&A, 355, 271

\bibitem[\protect\citeauthoryear{{Belloni}, {Motta}  \& {Mu{\~n}oz-Darias}}{{Belloni} et~al.}{2011}]{2011BASI...39..409B}
{Belloni} T.~M.,  {Motta} S.~E.,   {Mu{\~n}oz-Darias} T.,  2011, \mn@doi [Bulletin of the Astronomical Society of India] {10.48550/arXiv.1109.3388}, \href {https://ui.adsabs.harvard.edu/abs/2011BASI...39..409B} {39, 409}

\bibitem[\protect\citeauthoryear{Casella, Belloni  \& Stella}{Casella et~al.}{2005}]{Casella_2005}
Casella P.,  Belloni T.,   Stella L.,  2005, \mn@doi [The Astrophysical Journal] {10.1086/431174}, 629, 403

\bibitem[\protect\citeauthoryear{Chakrabarti \& Manickam}{Chakrabarti \& Manickam}{2000}]{Chakrabarti_2000}
Chakrabarti S.~K.,  Manickam S.~G.,  2000, \mn@doi [The Astrophysical Journal] {10.1086/312512}, 531, L41

\bibitem[\protect\citeauthoryear{Choudhury, Bhatt  \& Bhattacharyya}{Choudhury et~al.}{2015}]{Choudhury_2015}
Choudhury M.,  Bhatt N.,   Bhattacharyya S.,  2015, \mn@doi [Monthly Notices of the Royal Astronomical Society] {10.1093/mnras/stu2742}, 447, 3960–3972

\bibitem[\protect\citeauthoryear{Done, Gierlinski  \& Kubota}{Done et~al.}{2007}]{Done2007}
Done C.,  Gierlinski M.,   Kubota A.,  2007, \mn@doi [Astronomy and Astrophysics Review] {10.1007/s00159-007-0006-1}, 15, 1

\bibitem[\protect\citeauthoryear{Fan, Steiner, Bambi, Kara, Zhang  \& König}{Fan et~al.}{2025}]{Fan_2025}
Fan N.,  Steiner J.~F.,  Bambi C.,  Kara E.,  Zhang Y.,   König O.,  2025, \mn@doi [The Astrophysical Journal] {10.3847/1538-4357/adc25a}, 984, 31

\bibitem[\protect\citeauthoryear{Heil, Uttley  \& Klein-Wolt}{Heil et~al.}{2015}]{10.1093/mnras/stv191}
Heil L.~M.,  Uttley P.,   Klein-Wolt M.,  2015, \mn@doi [Monthly Notices of the Royal Astronomical Society] {10.1093/mnras/stv191}, 448, 3339

\bibitem[\protect\citeauthoryear{Homan \& Belloni}{Homan \& Belloni}{2005}]{Homan_2005}
Homan J.,  Belloni T.,  2005, \mn@doi [Astrophysics and Space Science] {10.1007/s10509-005-1197-4}, 300, 107–117

\bibitem[\protect\citeauthoryear{Ingram \& Done}{Ingram \& Done}{2012}]{Ingram_Done_2011}
Ingram A.,  Done C.,  2012, \mn@doi [Monthly Notices of the Royal Astronomical Society] {10.1111/j.1365-2966.2011.19885.x}, 419, 2369

\bibitem[\protect\citeauthoryear{Ingram \& Motta}{Ingram \& Motta}{2019}]{Ingram_2019}
Ingram A.~R.,  Motta S.~E.,  2019, \mn@doi [New Astronomy Reviews] {10.1016/j.newar.2020.101524}, 85, 101524

\bibitem[\protect\citeauthoryear{{Ingram} \& {van der Klis}}{{Ingram} \& {van der Klis}}{2015}]{2015MNRAS.446.3516I}
{Ingram} A.,  {van der Klis} M.,  2015, \mn@doi [MNRAS] {10.1093/mnras/stu2373}, \href {https://ui.adsabs.harvard.edu/abs/2015MNRAS.446.3516I} {446, 3516}

\bibitem[\protect\citeauthoryear{{Janiuk}, {Czerny}  \& {Siemiginowska}}{{Janiuk} et~al.}{2000}]{2000ApJ...542L..33J}
{Janiuk} A.,  {Czerny} B.,   {Siemiginowska} A.,  2000, \mn@doi [ApJl] {10.1086/312911}, \href {https://ui.adsabs.harvard.edu/abs/2000ApJ...542L..33J} {542, L33}

\bibitem[\protect\citeauthoryear{{Jones}, {Forman}, {Tananbaum}  \& {Turner}}{{Jones} et~al.}{1976}]{1976ApJ...210L...9J}
{Jones} C.,  {Forman} W.,  {Tananbaum} H.,   {Turner} M.~J.~L.,  1976, \mn@doi [ApJl] {10.1086/182291}, \href {https://ui.adsabs.harvard.edu/abs/1976ApJ...210L...9J} {210, L9}

\bibitem[\protect\citeauthoryear{{Katoch}, {Baby}, {Nandi}, {Agrawal}, {Antia}  \& {Mukerjee}}{{Katoch} et~al.}{2021}]{2021MNRAS.501.6123K}
{Katoch} T.,  {Baby} B.~E.,  {Nandi} A.,  {Agrawal} V.~K.,  {Antia} H.~M.,   {Mukerjee} K.,  2021, \mn@doi [MNRAS] {10.1093/mnras/staa3756}, \href {https://ui.adsabs.harvard.edu/abs/2021MNRAS.501.6123K} {501, 6123}

\bibitem[\protect\citeauthoryear{Lightman \& Eardley}{Lightman \& Eardley}{1974}]{lightman1974}
Lightman A.~P.,  Eardley D.~M.,  1974, \mn@doi [The Astrophysical Journal] {10.1086/181377}, 187, L1

\bibitem[\protect\citeauthoryear{Mishra, Fragile, Anderson, Blankenship, Li  \& Nalewajko}{Mishra et~al.}{2022}]{Mishra_2022}
Mishra B.,  Fragile P.~C.,  Anderson J.,  Blankenship A.,  Li H.,   Nalewajko K.,  2022, \mn@doi [The Astrophysical Journal] {10.3847/1538-4357/ac938b}, 939, 31

\bibitem[\protect\citeauthoryear{Nayakshin, Rappaport  \& Melia}{Nayakshin et~al.}{2000}]{Nayakshin2000}
Nayakshin S.,  Rappaport S.,   Melia F.,  2000, \mn@doi [The Astrophysical Journal] {10.1086/308876}, 535, 798

\bibitem[\protect\citeauthoryear{Neilsen, Remillard  \& Lee}{Neilsen et~al.}{2011}]{Neilsen_2011}
Neilsen J.,  Remillard R.~A.,   Lee J.~C.,  2011, \mn@doi [The Astrophysical Journal] {10.1088/0004-637X/737/2/69}, 737, 69

\bibitem[\protect\citeauthoryear{{Pahari} et~al.,}{{Pahari} et~al.}{2018}]{pahari}
{Pahari} M.,  et~al., 2018, \mn@doi [ApJ] {10.3847/1538-4357/aae53b}, \href {https://ui.adsabs.harvard.edu/abs/2018ApJ...867...86P} {867, 86}

\bibitem[\protect\citeauthoryear{Plant, Fender, Ponti, Muñoz-Darias  \& Coriat}{Plant et~al.}{2014}]{10.1093/mnras/stu867}
Plant D.~S.,  Fender R.~P.,  Ponti G.,  Muñoz-Darias T.,   Coriat M.,  2014, \mn@doi [Monthly Notices of the Royal Astronomical Society] {10.1093/mnras/stu867}, 442, 1767

\bibitem[\protect\citeauthoryear{{Rawat} et~al.,}{{Rawat} et~al.}{2019}]{2019ApJ...870....4R}
{Rawat} D.,  et~al., 2019, \mn@doi [ApJ] {10.3847/1538-4357/aaefed}, \href {https://ui.adsabs.harvard.edu/abs/2019ApJ...870....4R} {870, 4}

\bibitem[\protect\citeauthoryear{{Remillard}, {Morgan}, {McClintock}, {Bailyn}  \& {Orosz}}{{Remillard} et~al.}{1999}]{1999ApJ...522..397R}
{Remillard} R.~A.,  {Morgan} E.~H.,  {McClintock} J.~E.,  {Bailyn} C.~D.,   {Orosz} J.~A.,  1999, \mn@doi [ApJ] {10.1086/307606}, \href {https://ui.adsabs.harvard.edu/abs/1999ApJ...522..397R} {522, 397}

\bibitem[\protect\citeauthoryear{{Seifina}, {Titarchuk}  \& {Shaposhnikov}}{{Seifina} et~al.}{2014}]{sheifina}
{Seifina} E.,  {Titarchuk} L.,   {Shaposhnikov} N.,  2014, \mn@doi [ApJ] {10.1088/0004-637X/789/1/57}, \href {https://ui.adsabs.harvard.edu/abs/2014ApJ...789...57S} {789, 57}

\bibitem[\protect\citeauthoryear{{S{\k{a}}dowski}, {Wielgus}, {Narayan}, {Abarca}, {McKinney}  \& {Chael}}{{S{\k{a}}dowski} et~al.}{2017}]{2017MNRAS.466..705S}
{S{\k{a}}dowski} A.,  {Wielgus} M.,  {Narayan} R.,  {Abarca} D.,  {McKinney} J.~C.,   {Chael} A.,  2017, \mn@doi [MNRAS] {10.1093/mnras/stw3116}, \href {https://ui.adsabs.harvard.edu/abs/2017MNRAS.466..705S} {466, 705}

\bibitem[\protect\citeauthoryear{{Stella} \& {Vietri}}{{Stella} \& {Vietri}}{1998a}]{stella_luigi}
{Stella} L.,  {Vietri} M.,  1998a, \mn@doi [ApJl] {10.1086/311075}, \href {https://ui.adsabs.harvard.edu/abs/1998ApJ...492L..59S} {492, L59}

\bibitem[\protect\citeauthoryear{{Stella} \& {Vietri}}{{Stella} \& {Vietri}}{1998b}]{1998ApJ...492L..59S}
{Stella} L.,  {Vietri} M.,  1998b, \mn@doi [ApJl] {10.1086/311075}, \href {https://ui.adsabs.harvard.edu/abs/1998ApJ...492L..59S} {492, L59}

\bibitem[\protect\citeauthoryear{{Tagger} \& {Pellat}}{{Tagger} \& {Pellat}}{1999}]{1999A&A...349.1003T}
{Tagger} M.,  {Pellat} R.,  1999, \mn@doi [AA] {10.48550/arXiv.astro-ph/9907267}, \href {https://ui.adsabs.harvard.edu/abs/1999A&A...349.1003T} {349, 1003}

\bibitem[\protect\citeauthoryear{Trudolyubov}{Trudolyubov}{2001}]{Trudolyubov2001}
Trudolyubov S. e.~a.,  2001, \mn@doi [The Astrophysical Journal] {10.1086/321375}, 554, 383

\bibitem[\protect\citeauthoryear{{Trudolyubov}, {Borozdin}  \& {Priedhorsky}}{{Trudolyubov} et~al.}{2001}]{2001MNRAS.322..309T}
{Trudolyubov} S.~P.,  {Borozdin} K.~N.,   {Priedhorsky} W.~C.,  2001, \mn@doi [MNRAS] {10.1046/j.1365-8711.2001.04073.x}, \href {https://ui.adsabs.harvard.edu/abs/2001MNRAS.322..309T} {322, 309}

\bibitem[\protect\citeauthoryear{{Varni{\`e}re}, {Rodriguez}  \& {Tagger}}{{Varni{\`e}re} et~al.}{2002}]{2002A&A...387..497V}
{Varni{\`e}re} P.,  {Rodriguez} J.,   {Tagger} M.,  2002, \mn@doi [AA] {10.1051/0004-6361:20020401}, \href {https://ui.adsabs.harvard.edu/abs/2002A&A...387..497V} {387, 497}

\bibitem[\protect\citeauthoryear{Wagoner, Silbergleit  \& Ortega-Rodriguez}{Wagoner et~al.}{2001}]{Wagoner:2001uj}
Wagoner R.~V.,  Silbergleit A.~S.,   Ortega-Rodriguez M.,  2001, \mn@doi [Astrophys. J. Lett.] {10.1086/323655}, 559, L25

\bibitem[\protect\citeauthoryear{Yang \& et al.}{Yang \& et~al.}{2022}]{Yang_2022}
Yang Z.-x.,  et al. 2022, \mn@doi [The Astrophysical Journal] {10.3847/1538-4357/ac84d6}, 937, 33

\bibitem[\protect\citeauthoryear{{Zhao} et~al.,}{{Zhao} et~al.}{2023}]{Zhao}
{Zhao} Q.~C.,  et~al., 2023, \mn@doi [MNRAS] {10.1093/mnras/stad1965}, \href {https://ui.adsabs.harvard.edu/abs/2023MNRAS.524.3215Z} {524, 3215}

\bibitem[\protect\citeauthoryear{van~der Klis}{van~der Klis}{1989}]{vanderKlis1989}
van~der Klis M.,  1989, Fourier Techniques in X-Ray Timing.
Springer Netherlands, Dordrecht, pp 27--69, \mn@doi{10.1007/978-94-009-2273-0_3}, \url {https://doi.org/10.1007/978-94-009-2273-0_3}

\bibitem[\protect\citeauthoryear{{van der Klis}}{{van der Klis}}{1994}]{1994ApJS...92..511V}
{van der Klis} M.,  1994, \mn@doi [ApJs] {10.1086/192006}, \href {https://ui.adsabs.harvard.edu/abs/1994ApJS...92..511V} {92, 511}

\makeatother
\end{thebibliography}

\end{document}